# The VMC ESO Public Survey

Maria-Rosa L. Cioni[1,2]
Peter Anders[3]
Gemma Bagheri[1]
Kenji Bekki[4]
Gisella Clementini[5]
Jim Emerson[6]
Chris J. Evans[7]
Bi-Qing For[4]
Richard de Grijs[8]
Brad Gibson[9]
Léo Girardi[10]
Martin A. T. Groenewegen[11]
Roald Guandalini[12]
Marco Gullieuszik[10]
Valentin D. Ivanov[13]
Devika Kamath[12]
Marcella Marconi[14]
Jean-Baptiste Marquette[15]
Brent Miszalski[16]
Ben Moore[17]
Maria Ida Moretti[14]
Tatiana Muraveva[5]
Ralf Napiwotzki[1]
Joana M. Oliveira[18]
Andrés E. Piatti[19]
Vincenzo Ripepi[14]
Krista Romita[20]
Stefano Rubele[10]
Richard Sturm[21]
Ben Tatton[18]
Jacco Th. van Loon[18]
Mark I. Wilkinson[22]
Peter R. Wood[23]
Simone Zaggia[10]

[1] University of Hertfordshire, United Kingdom
[2] Leibniz-Institut für Astrophysik Potsdam, Germany
[3] National Astronomical Observatory of China, China
[4] ICRAR, University of Western Australia, Australia
[5] INAF, Osservatorio Astronomico di Bologna, Italy
[6] Queen Mary University London, United Kingdom
[7] Astronomy Technology Centre, Edinburgh, United Kingdom
[8] Peking University, China
[9] University of Central Lancashire, United Kingdom
[10] INAF, Osservatorio Astronomico di Padova, Italy
[11] Royal Observatory of Belgium, Belgium
[12] Institute of Astronomy, KU Leuven, Belgium
[13] ESO
[14] INAF, Osservatorio Astronomico di Capodimonte, Italy
[15] Institut d'Astrophysique de Paris, France
[16] South African Astronomical Observatory, Cape Town, South Africa
[17] University of Zurich, Switzerland
[18] Lennard-Jones Laboratories, Keele University, United Kingdom
[19] Observatorio Astronómico, Universidad National de Córdoba, Argentina
[20] University of Florida, USA
[21] Max-Planck-Institut für extraterrestrische Physik, Germany
[22] University of Leicester, United Kingdom
[23] Australian National University, Australia

The VISTA near-infrared $YJKs$ survey of the Magellanic Clouds system (VMC) has entered its core phase: about 50 % of the observations across the Large and Small Magellanic Clouds (LMC, SMC), the Magellanic Bridge and Stream have already been secured and the data are processed and analysed regularly. The initial analyses, concentrated on the first two completed tiles in the LMC (including 30 Doradus and the South Ecliptic Pole), show the superior quality of the data. The photometric depth of the VMC survey allows the derivation of the star formation history (SFH) with unprecedented quality compared to previous wide-area surveys, while reddening maps of high angular resolution are constructed using red clump stars. The multi-epoch $Ks$-band data reveal tight period–luminosity relations for variable stars and permit the measurement of accurate proper motions of the stellar populations. The VMC survey continues to acquire data that will address many issues in the field of star and galaxy evolution.

## The VMC survey

The VMC survey observations are obtained with the infrared camera VIRCAM mounted on VISTA and reach stars down to a limiting magnitude of ~ 22 (5σ Vega) in the $YJKs$ filters. The VMC strategy involves repeated observations of tiles across the Magellanic system, where one tile covers approximately uniformly an area of ~ 1.5 square degrees in a given band with three epochs at $Y$ and $J$, and 12 epochs at $Ks$ spread over a time range of one year or longer. Individual $Ks$ epochs refer each to exposure times of 750 s and reach a limiting magnitude of ~ 19 for sources with photometric errors < 0.1 mag. The VMC data are acquired under homogeneous sky conditions, since observations take place in service mode, and their average quality corresponds to a full width at half maximum < 1 arcsecond. The VISTA astrometry, which is based on 2MASS, results in positional accuracies within 25 milliarcseconds (mas) across a tile. The VMC data are reduced with the VISTA Data Flow System (VDFS) pipeline and are archived both at the VISTA Science Archive and at ESO. Further details about the VMC survey[1] are given in Cioni et al. (2011).

## Stellar populations

One of the main goals of the VMC survey is the identification and characterisation of the mixture of stellar populations that have made up the Magellanic system over time. The star formation history of field stars, the physical parameters of stellar clusters, the links between these and the structure and dynamical processes are all embedded in the VMC data. Extracting a comprehensive picture of the system represents our major challenge, but fortunately we have access to sophisticated tools with which to do the job. In Rubele et al. (2012), we demonstrated that by using two colour–magnitude diagrams (CMDs) simultaneously, and a grid of models at various ages and metallicities, we could derive spatially resolved SFHs where systematic errors in the star formation rate and age–metallicity relations are reduced by a factor of two, relative to previous work, after accounting for the geometry of the galaxy. In our study we independently derive the mean extinction and distance modulus for twelve subsections of the original tiles.

In Figure 1 we show the CMD of a tile in the SMC including the Milky Way (MW) globular cluster 47 Tuc, highlighting the complexity of the SFH analysis in decomposing the different stellar populations. Using custom-derived point spread function photometry, we can push the sensitivity of the VMC data to highly crowded regions. Together with the wide area covered by VMC we will be able to investigate not only substructures in the LMC and SMC, but also streams attached to the 47 Tuc cluster, for example, as well as detecting the members of hundreds of stellar clusters in the Magellanic system waiting to be characterised.

## The reddening map of the 30 Doradus field

Dust causes uncertainties in the measurements of the SFH and the structure of galaxies. Red clump stars (0.8–2 $M_\odot$ and 1–10 Gyr old) are useful probes of interstellar reddening because of their large number and relatively fixed luminosity. Red clump stars belonging to the tile LMC 6_6 are selected from their location in the ($J–Ks$) vs. $Ks$ CMD. Then, the amount of total reddening (along the line of sight and within the LMC) in terms of colour excess is obtained for each of ~ 150 000 stars with respect to its intrinsic colour. The latter is derived accordingly from stellar evolution models, accounting for variations with age and metallicity. Extinction is subsequently converted into hydrogen gas column density. Compared to reddening maps produced using the same method at optical wavelengths, the near-infrared VMC data are more sensitive to higher extinction. Compared to H I observations we derive that, on average, half of the stars lie in front of the H I column and hydrogen becomes molecular in the dustiest clouds; the transition begins at $N_{HI} \cong 4 \times 10^{21}$ cm$^{-2}$. Figure 2 shows the location of molecular clouds superposed on the distribution of hydrogen column density inferred from the VMC data. There is overall agreement with maps of dust emission at 24 µm and 70 µm (see Tatton et al., 2013). Reddening maps will be created for other tiles in the VMC survey allowing red clump stars to be de-reddened; these results will be used in calculating the three-dimensional (3D) structure of the Magellanic system.

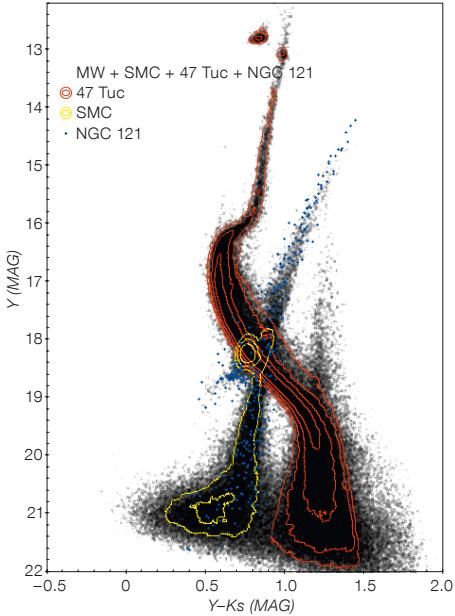

Figure 1. Colour–magnitude diagram of stellar sources in tile SMC 5_2. All sources are shown in grey; stars belonging to the SMC cluster NGC 121 are indicated in blue; the field population of the SMC is indicated with yellow contours; and the stars of the Milky Way cluster 47 Tuc are shown with red contours.

### Variable stars

The other main goal of the VMC survey is the measurement of the 3D structure of the Magellanic system. Classical Cepheids are primary distance indicators and in the near-infrared obey period–luminosity (PL) relations that are less affected by reddening, chemical composition and nonlinearity than those at optical wavelengths, resulting in smaller intrinsic dispersion. First results for classical Cepheids in the tiles LMC 6_6 (Figure 3, left) and 8_8 have been presented in Ripepi et al. (2012). The identification of the variables is derived from the EROS-2 and OGLE-III catalogues and their VMC $K_s$ light curves are very well sampled, with at least 12 epochs, and high precision, with typical errors of 0.01 mag, or better, for individual phase points. The $K_s$ mag of the faintest Cepheids in the LMC, which are mostly first overtone pulsators, was measured for the first time thanks to the VMC observing strategy. Photometry for the brightest fundamental mode Cepheids (periods > 23 days), exceeding the linearity regime of VMC data, are taken from the literature. The dispersion of the PL relations is ~ 0.07 mag.

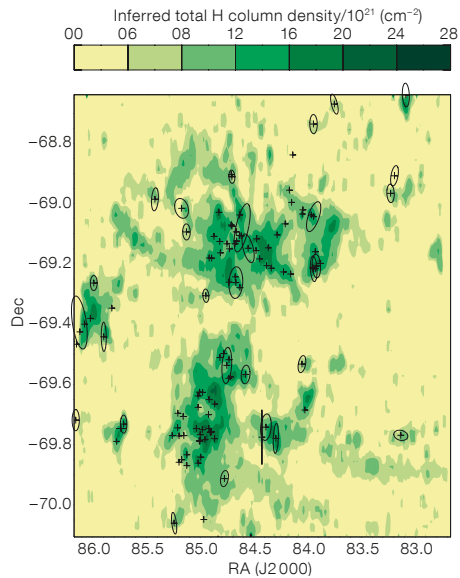

Figure 2. Hydrogen column density map inferred from VMC data in tile LMC 6_6 identifying regions where $N_{HI} > 8 \times 10^{21}$ cm$^{-2}$. Crosses represent molecular clouds catalogued in the literature and ellipses highlight those with measured properties.

Anomalous Cepheids (1.3–2.1 M$_\odot$ and [Fe/H] ≈ –1.7 dex) also play an important role both as distance indicators and stellar population tracers. The VMC survey has already observed many of the anomalous Cepheids discovered by the OGLE project in the LMC. These stars obey a tight PL relation in the $K_s$-band with a dispersion of 0.10 mag (Figure 3, right) that is shown for the first time in Ripepi et al. (2013).

Cepheids (< 200 Myr old) are mainly concentrated towards the bar and in a northwest spiral arm of the LMC as well as in the central region of the SMC. Eclipsing binaries composed of main sequence stars trace a similar distribution, but with clustering mainly occurring in regions of recent star formation. On the other hand, RR Lyrae variable stars (> 10 Gyr old) are smoothly distributed and likely trace the haloes of the galaxies. These stars also follow a PL relation that is tight in the $K_s$-band. The VMC properties and the strategy to measure distances and infer the system 3D geometry of different age components from the variable stars is described in Moretti et al. (2013).

The magnitude of the brightest VMC objects (10 < $K_s$ < 12), which may be saturated in their central regions, is well recovered by the VDFS pipeline by integrating the flux in the outer parts. Most of these sources are asymptotic giant branch (AGB) stars. By fitting spectral energy distributions, created from the combination of VMC data and data at other wavelengths, with dust radiative transfer models, it is possible to derive mass-loss rates, luminosities and spectral classifications that offer strong constraints on AGB evolutionary models (Gullieuszik et al., 2012). These variable stars obey PL relations that may also be useful as distance and structure indicators.

### The proper motion of the LMC

The astrometric accuracy and the photometric sensitivity of observations made with VISTA are of sufficient quality to select a large sample of targets and measure their proper motion. The proper motion of the LMC is measured from the combination of 2MASS and VMC data that span a time range of ~ 10 years and from VMC data alone across a time baseline of ~ 1 year (Cioni et al., 2013b). Different types of LMC stars (e.g., red giant branch, red clump and main sequence stars, as well as variable stars) are selected from their location in the ($J$–$K_s$) vs. $K_s$ CMD, and from lists of known objects, where MW foreground stars and background galaxies are also easily distinguished (Figure 4, left). The proper motion of ~ 40 000 LMC stars in the tile, with respect to ~ 8000 background galaxies, is $\mu_\alpha \cos(\delta)$ = +2.20 ± 0.06 mas yr$^{-1}$ and $\mu_\delta$ = 1.70 ± 0.06 mas yr$^{-1}$. This value is in excellent agreement with previous ground-based measurements but our statistical uncertainties are a factor of three smaller and are directly comparable to uncertainties derived with the Hubble Space Telescope. The error budget is at present dominated by systematic uncertainties (a few mas yr$^{-1}$), but these will decrease due to the improved reduction of the VISTA data and the increase in the time baseline.

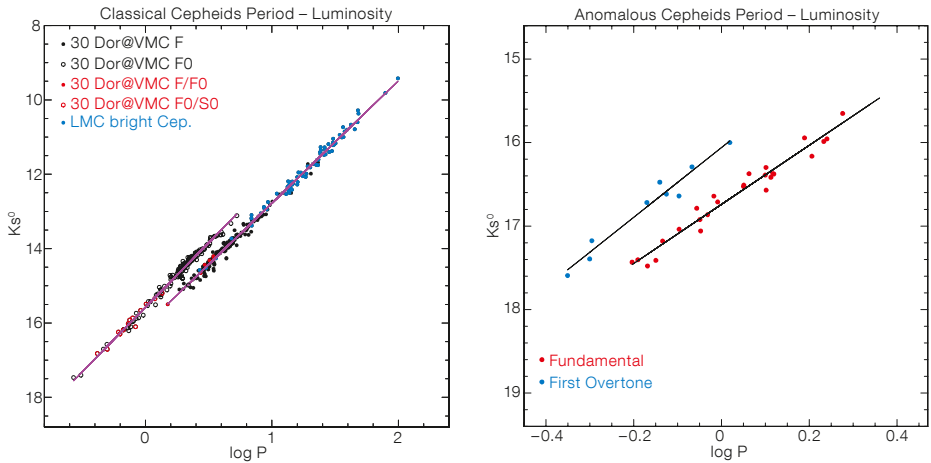

Figure 3. $Ks$-band period–luminosity relation for classical Cepheids in the 30 Dor field (left) and for LMC anomalous Cepheids (right). Fundamental and first overtone pulsators are indicated in blue and red respectively, and the solid lines are the result of least squares fits to the data.

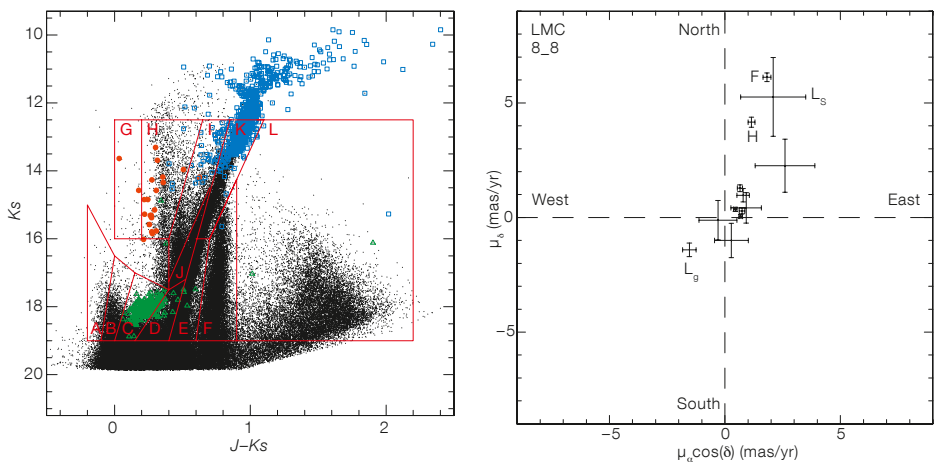

Figure 4. Left: Colour–magnitude diagram (CMD) of VMC sources in tile LMC 8_8. Region boundaries distinguish stars of different types. Colours highlight Cepheids (red), long period variables (blue) and RR Lyrae stars (green). Right: Proper motion derived from stars in each CMD region. Points corresponding to Milky Way stars (F, H and $L_s$) and to background galaxies ($L_g$) are indicated, while, for clarity, those corresponding to LMC stars are not labelled.

The mean proper motion of stars in each CMD region is shown in Figure 4 (right). The proper motions of LMC stars are clustered around zero except for region A, despite the large uncertainty. The proper motion of MW stars (regions F and H) is clearly distinct from those of LMC stars and background galaxies (region $L_g$). The latter refers to a sub-group of objects in region L with a galaxy-like morphology. Those with a stellar-like morphology are likely late-type MW dwarfs, since their proper motion value ($L_s$) is consistent with that of other MW stars. Using the best fitting model from Rubele et al. (2012) we can associate an age to each CMD region. We then measure a decrease of the proper motion with age where young stars stretch out to the northeast and old stars to the southwest. This difference is linked to both kinematic differences between young and old stars and to different hosting structures.

An alternative and established reference system is made of background quasars and hundreds of them have already been found behind the central regions of the Magellanic Clouds as a result of spectroscopic observations of candidates identified from the OGLE light curves. The combination of VMC colours and $Ks$-band variability is also a valuable method to identify candidate quasars (Cioni et al., 2013a). The contamination by background galaxies, foreground late-type dwarf stars, young stellar objects and planetary nebulae, for which some nebular morphologies are revealed for the first time with VMC data (Miszalski et al., 2011), is reduced to ~ 20 %.

## Perspective

The first years of observations have shown that the VMC data meet expectations and our understanding of the SFH across the Magellanic system will no longer be limited to small area observations. With the approaching completion of the SMC and Bridge areas it will be possible to start comparing theoretical predictions with observations on the age and metallicity distributions and their relations with the interaction between the LMC and SMC, the formation of the Bridge and the existence of stripped stars. The VMC survey has a high legacy value and represents the sole counterpart in the $Ks$-band to current and future ground-based (STEP at the VST, SkyMapper, SMASH at the Blanco 4-metre, Large Synoptic Survey Telescope [LSST]) and space-based imaging missions (e.g., Gaia and Euclid) targeting or including the Magellanic system. It also provides a wealth of targets for wide-field spectroscopic follow-up investigations, e.g., with the Apache Point Observatory Galactic Evolution Experiment (APOGEE)-South, High Efficiency and Resolution Multi-Element Spectrograph (HERMES) at the Anglo-Australian Observatory, 4-metre Multi-Object Spectroscopic Telescope (4MOST) and Multi-Object Optical and Near-infrared Spectrograph (MOONS) at ESO.

### References

Cioni, M.-R. L. et al. 2011, A&A, 527, A116, Paper I
Cioni, M.-R. L. et al. 2013a, A&A, 549, A29, Paper VI
Cioni, M.-R. L. et al. 2013b, A&A, accepted, Paper IX
Gullieuszik, M. et al. 2012, A&A, 537, A105, Paper III
Miszalski, B. et al. 2011, A&A, 531, A157, Paper II
Moretti, M. I. et al. 2013, A&A, accepted, Paper X
Ripepi, V. et al. 2012, MNRAS, 424, 1807, Paper V
Ripepi, V. et al. 2013, MNRAS, accepted, Paper VIII
Rubele, S. et al. 2012, A&A, 537, A106, Paper IV
Tatton, B. L. et al. 2013, A&A, 554, A33, Paper VII

### Links

[1] VMC survey: http://star.herts.ac.uk/~mcioni/vmc/